\documentstyle[preprint,aps]{revtex}

\tighten
\begin{document}

\draft

\title{
         GCM study of hexadecapole correlations in
         superdeformed $^{194}$Hg.
}

\author     {
             P. Magierski\thanks{Permanent address:
             Institute of Physics, Warsaw University of Technology,
                       ul. Koszykowa 75, PL--00-662 Warsaw, Poland }
             and
             P.-H. Heenen\thanks{Directeur de Recherches FNRS.} \\
             {\em Service de Physique Nucl\'{e}aire Th\'{e}orique,} \\
             {\em U.L.B - C.P. 229, B 1050 Brussels, Belgium} \\
and        \\
  W. Nazarewicz \\
       {\em Department of Physics and Astronomy, University of  Tennessee,}\\
       {\em Knoxville, Tennessee 37996, U.S.A.}\\
       {\em Physics Division, Oak Ridge National Laboratory}\\
       {\em P. O. Box 2008, Oak Ridge, Tennessee 37831, U.S.A.}
            }% end author

\maketitle

\begin{abstract}
The role of hexadecapole correlations in the lowest superdeformed
band of $^{194}$Hg is studied by self consistent mean field
methods.
The generator coordinate method with particle number projection
has been applied using Hartree-Fock wave functions
defined along three different hexadecapole paths. In all cases,
the ground state is not significantly affected by hexadecapole
correlations and the energies of the
corresponding first excited hexadecapole
vibrational states lie high in energy. The effect of
rotation is investigated with the
Skyrme-Hartree-Fock-Bogolyubov method and a
zero range density-dependent
pairing interaction.
\end{abstract}

\pacs{PACS numbers: 21.10.Ky, 21.10.Re,  21.60.-n, 27.80.+w}

\narrowtext

\section{INTRODUCTION}
Recent experimental data on the spectra of superdeformed (SD) bands
in $^{149}$Gd \cite{Fli93}, $^{153}$Dy \cite{Ced},
$^{194}$Hg \cite{Ced94}, and $^{131,132}$Ce \cite{Sem95}
 have put into evidence
a $\Delta I=4$ staggering of the dynamical moment of inertia
of these bands.
It manifests itself
in systematic shifts of the energy levels which are alternately
pushed down and up with respect to a purely rotational sequence.
The amplitude of this staggering is of the order of 50 eV.
Since these oscillations
distinguish states  differing by four units of angular momentum,
it seems natural to explain their origin
by a coupling between the
rotational motion and hexadecapole vibrations~\cite{Fli93,Pek83}.
This scenario requires the presence in the mean field
of the nucleus of
a small component with a four-fold rotational symmetry,
i.e.,  a symmetry associated with a $\displaystyle{\frac{\pi}{2}}$
rotation around one of the principal axes of the
quadrupole tensor.
%nuclear density.
The strongly prolate-deformed intrinsic Hamiltonian
is then slightly perturbed
by a term with the C$_4$ symmetry.

The effect of such terms
has been investigated in phenomenological models
with the C$_4$ symmetry axis coinciding either with the symmetry axis
($z$-axis) of the
quadrupole tensor \cite{HM94,MB94,Mac95} or with the
C$_4$ perturbation quantized along the rotation axis ($x$-axis)
 \cite{Fli93,Pav94}.
Hamamoto and Mottelson \cite{HM94} have studied the properties of
a quartic rotational Hamiltonian. Staggering appears then as a
result of tunneling between the four
equivalent minima of the total energy surface
of the Hamiltonian due to the $K$-mixing.
Burzy\'nski {\it et al.}~\cite{BM94} have obtained the
exact solutions for a system of $N$ identical particles
in a single $j$-shell interacting through a
multipole-multipole  Hamiltonian with  quadrupole and
hexadecapole terms. They  have shown
that the staggering in  $J^{(2)}$ may occur in certain cases.

All these studies have assumed the presence of a hexadecapole
term in the Hamiltonian of the SD nucleus.
It still remains to explain the origin of such a term.
Ragnarsson~\cite{Rag94} has considered triaxial hexadecapole
deformations in a study at high spins of SD bands in the $A$=150
mass region using the shell correction approach with a Nilsson
potential.
The effect of $\epsilon_{44}$
deformation has
been found extremely small.
Frauendorf {\it et al.} \cite{Fra94} have applied the tilted axis cranking
formalism also to a Nilsson Hamiltonian with a hexadecapole field,
$Y_{44}+Y_{4-4}$, and have concluded that it is impossible to generate
a staggering of the observed order of magnitude
even assuming
$\epsilon_{44}$ values
as large as 0.1.

In the present work, based
on self-consistent  methods with
Skyrme interactions,
the importance of hexadecapole correlations in
the lowest SD band of $^{194}$Hg is investigated.
Firstly, we analyze
 dynamical
hexadecapole correlations without  rotation
by means of the generator coordinate
method (GCM) with particle number projection \cite{HBD93}.
Secondly,
cranking calculations have been
performed within the Hartree-Fock-Bogoliubov +
Lipkin Nogami (HFBLN) method  \cite{Mag93,GBD94} with
a zero-range density dependent interaction \cite{Ter95}
in the pairing channel.

\section{THE METHODS}

\subsection{The Generator Coordinate method}
The GCM formalism used in this study has been presented in detail
in Ref. \cite{HBD93}. It permits one
to study the dynamics of the nucleus
as a function of one or several collective coordinates
using generating  functions with good particle number.
The intrinsic wave functions, ${\vert\Phi\rangle}$, are obtained
by constrained HF+BCS calculations with the Routhian
\begin{equation}\label{EP}
E' = {\langle\Phi\vert} {\hat H} {\vert\Phi\rangle}-
\lambda{\langle\Phi\vert}{\hat N}{\vert\Phi\rangle}-\lambda_2
{\langle{\Delta\hat{N}}^2\rangle}
      -q_{44}^{i}({\langle\Phi\vert}
\hat{Q}_{44}^{i}{\vert\Phi\rangle} - Q_{44}^{i})^{2}. \label{ener}
\end{equation}
In Eq.~(\ref{EP})
${\hat N}$ is the particle number operator and
$Q_{44}^{i}$ denotes the hexadecapole moment quantized
along the $i$ axis, i.e.,
\begin{equation}\label{Y44}
Q_{44}^{i}=\frac{1}{2}\int r^{4}(Y_{44}^{i} + Y_{4-4}^{i})\rho({\bf r})d^{3}r,
\end{equation}
where $i=x,\,z$ and $\rho$ denotes the
selfconsistenf HFBCS density distribution. The functions
$Y_{44}^{x}$ and $Y_{44}^{z}$ are
the spherical harmonics along the $x$ and $z$ axis respectively.
Thus $Q_{44}^{i}$ has
a fourfold (C$_4$) symmetry  along the $i$th axis.
 The energy
dependence on  hexadecapole distortions
around the SD minimum of $^{194}$Hg
is determined using constraints on either $Q_{44}^{z}$ or $Q_{44}^{x}$.
We have chosen $z$ as the symmetry axis of the quadrupole tensor,
whereas $x$ denotes the  perpendicular
 axis.

In the particle-hole channel, we employ
the Skyrme interaction SkM$^*$ \cite{Bar82}.
In the particle-particle (pairing) channel,
 the state-independent  seniority interation is used.
An approximate variation after projection
on particle number is performed by means of the
Lipkin-Nogami method \cite{PRA73}.

In this study it is assumed that the density
distribution has three symmetry planes. The HF
equations have then been solved on a rectangular mesh
$12\times 12\times 16$\,fm$^{3}$, corresponding to the
octant with
positive values
of the three cartesian coordinates.
Single-particle wave functions
are discretized with a mesh size of 1 fm.
 We have checked that the results
are not qualitatively modified when increasing the size of the box
or changing the mesh size.

 From the HFBCS wave functions, states
$|NZ\rangle$ with the correct number of particles are obtained by
projection:
\begin{equation}
|NZ\rangle = \hat{P}_{NZ}{\vert\Phi\rangle} \label{prwave},
\end{equation}
where
\begin{equation}
\hat{P}_{NZ}=\frac{1}{\pi}\int_{0}^{\pi} \exp\left\{i\phi_{N}
(\hat{N}-N)\right\}d\phi_{N}
             \frac{1}{\pi}\int_{0}^{\pi} \exp\left\{i\phi_{Z}
(\hat{Z}-Z)\right\}d\phi_{Z}
\end{equation}
is the
projection operator on neutron and proton numbers.
The integrations over the gauge angles
are approximated by  $n$-point trapeze formulae.
The choice $n=7$ ensures that the particle number dispersion
never exceeds 0.001.

The GCM wave function is a superposition of projected wave functions
corresponding to different hexadecapole deformations:
\begin{equation}
{\vert\Phi\rangle}_{\rm GCM} = \int f(Q_{44}^{i}) |NZ; Q_{44}^{i} \rangle
dQ_{44}^{i}
\end{equation}
The weight function  $f$  is obtained from the
Hill-Wheeler equation:
\begin{equation}
\int \left[\langle NZ;Q_{44}^{i}|\hat{H}|NZ; Q_{44}^{i'}\rangle
- E \langle NZ; Q_{44}^{i}|NZ; Q_{44}^{i'}\rangle \right]f(Q_{44}^{i'})
  dQ_{44}^{i'} = 0
\end{equation}
where $\hat{H}$ is the many-body Skyrme+pairing Hamiltonian.
Finally, the  collective
wave functions, $g$,  can be obtained by an integral transformation
of  $f(Q_{44}^{i})$:
\begin{equation}\label{coll}
g(Q_{44}^{i})=\int \langle NZ; Q_{44}^{i}|NZ; Q_{44}^{i'}
\rangle^{\frac{1}{2}}
              f(Q_{44}^{i'})dQ_{44}^{i'}.
\end{equation}

\subsection{The cranked HFBLN method}

The HFBLN method
for rotating nuclei has been presented in Refs. \cite{Mag93,GBD94}.
As in the GCM calculations without rotation, the
SkM$^{*}$ parametrization of the Skyrme force
has been used in the particle-hole channel.
In the pairing channel, the density-dependent pairing interaction,
\begin{equation}\label{e2}
V_P=\frac{V_0}{2}(1-P_\sigma)
\delta({\vec r}_1-{\vec r}_2)
\left[1-\alpha\rho\left(\frac{{\vec r}_1+{\vec r}_2}{2}\right)\right],
\end{equation}
has been employed.
The strength $V_0$ has been chosen equal to 880 MeV  and
the reference density, $\rho_c=\alpha^{-1}$, to 0.16 fm$^{-3}$.
It has been shown recently \cite{Ter95}
that the introduction of a
zero-range density dependent pairing interaction
improves considerably the agreement between the HFBLN
calculation and the experimental data for SD bands in the $A$=190
mass region.
In particular, the experimental
transition energies in the SD band of $^{194}$Hg
are reproduced with a
systematic error lower than 10 keV over 12 transitions.

Hexadecapole constraints have been introduced in the
cranked  HFBLN equations in the same way as in the
nonrotating case.

\section{RESULTS}

 The GCM collective path is determined
 by a full minimization of the energy
with constraints on the hexadecapole moment alone,
as defined by Eq.~(\ref{ener}).
This does not guarantee that
other collective variables do not change
along the path, i.e., that the hexadecapole mode
is decoupled from all the other modes.
In the case of such a coupling, one should in principle
enlarge the collective subspace
and include  other variables
as generator coordinates.

In our GCM calculations with the $Q_{44}^z$ constraint,  the
quadrupole-hexadecapole coupling is very weak. Namely,
the mass quadrupole
moment, $Q_{20}$=$Q_{20}^z$=$r^2 Y_{20}^z$,
does not vary by more than $1b$ in the whole considered range
of $Q_{44}^z$ values.
This is not true for
the $Q_{44}^x$ trajectory.
Indeed,
the $Q_{44}^x$ moment can be expressed in terms of
 hexadecapole moments $Q_{4\mu}^z$ quantized along the $z$-axis:
\begin{equation}\label{expansion}
Q_{44}^x = \frac{1}{8}Q_{44}^z + \frac{\sqrt{7}}{4}Q_{42}^z
+ \frac{\sqrt{70}}{16}Q_{40}^z.
\end{equation}
Consequently,   $Q_{44}^{x}$
 has $\mu$=0 and 2 components along the $z$ axis
which are strongly coupled to the quadrupole moments
$Q_{20}^z$ and $Q_{22}^z$.
To avoid many-dimensional GCM calculations in the $Q_{44}^{x}$
direction,
we have defined two different trajectories.
In the first version, the quadrupole moment
varies self-consistently as a function of $Q_{44}^{x}$.
In the second version, the quadrupole moment is constrained
to its value at the SD minimum.

The potential energy curves corresponding to
projected wave functions (\ref{prwave})
are plotted in Fig. \ref{FIG1}
as  functions of $Q_{44}^{z}$ (top) and $Q_{44}^{x}$ (bottom).
The evolution of the
quadrupole moment $Q_{20}^z$ as a function of $Q_{44}^{x}$
is plotted on Fig.~\ref{FIG2}. In
the same figures are shown the results
of the HFBLN calculation performed at $I$=42.

According to the calculations,
the collective motions along the $Q_{44}^{x}$ and $Q_{44}^{z}$
paths are decoupled.
Namely, in the $Q_{44}^z$ version of the calculations,
 the value of $Q_{44}^{x}$
does not vary by more than 1000\,fm$^4$ along the collective path.
Also in the $Q_{44}^{x}$ version
the value of $Q_{44}^{z}$ remains always compatible
with zero.

The potential energy
curves shown in Fig.~\ref{FIG1} exhibit
similar dependences as  functions of the hexadecapole constraints.
This means that,  qualitatively, the response of the nucleus to the
hexadecapole perturbation of the average field is the same in both cases.
As expected, relaxing the constraint on the quadrupole moment
(solid line in Fig.~\ref{FIG1})
leads to a slight decrease of the energy. In spite of the fact that
the quadrupole moment shows large variation along
the unconstrained path (see Fig.~\ref{FIG2}),
 the corresponding gain in energy never exceeds
a few hundred keV.

The effect of rotation on total potential energies
seems to be fairly
weak, although the conditions of the calculations
with or without rotation are not exactly the same.
The curves are surprisingly close in the $Q_{44}^z$ variant.
In the calculations with a constraint
along the rotation axis, there is a slight shift
of the minimum toward smaller
hexadecapole moments at $I$=42 but the hexadecapole stiffness is very
similar.
The small shift in the equilibrium value of $Q_{44}^x$ with
angular momentum can be attributed to the rotation-induced
shape change in the SD yrast band of $^{194}$Hg. As discussed
in Ref.~\cite{Sat91}, a systematic decrease in
both a $Q_{20}^z$ and $Q_{40}^z$ with rotational frequency
is expected for SD yrast band in $^{194}$Hg. According
to the self-consistent calculations, the expectation value of
$Q_{42}^z$ is almost zero
(i.e., at least two orders of magnitude smaller
than $Q_{40}^z$). Consequently, according to Eq~(\ref{expansion}),
the equilibrium value of $Q_{44}^x$ should follow closely
the value of $Q_{40}^z$. Indeed,
the shift in $Q_{40}^z$ from approximately 38000 ($I$=0)
 to 35300 ($I$=42) fm$^4$
is consistent with the results  displayed in Fig.~\ref{FIG1}.

It is worth noting that the quadrupole moments minimizing the energy
for small values of $Q_{44}^{x}$ are  rather
different for $I$=0 and $I$=42 (see Fig.~\ref{FIG2}).
This is probably more
 related to the different pairing interactions employed
and to particle number
projection than to rotation.
Indeed, the quadrupole moments
of the unprojected HFBCS states ${\vert\Phi\rangle}$
with and without rotation
vary in a similar manner as functions of $Q_{44}^x$.
On the other hand, the neutron pairing
energy which is of the order of 4.0 MeV
for both pairing interactions around the SD minimum,
increases up to 6.5 MeV for a  seniority
interaction  at small $Q_{44}^{x}$  values.
This leads
to  large differences between the
projected and unprojected wave functions.

A static calculation does not give access to a scale parameter
which would permit one to compare directly the evolution of
the energy as a function of $Q_{44}^{x}$ and $Q_{44}^{z}$.
As discussed above, the large values of
$Q_{44}^{x}$ are related to the
large
hexadecapole moment
of the nucleus along the $z$ axis, while
the mean value of $Q_{44}^{z}$
at the SD minimum is compatible with $0.$ A first
crude measure of the variation of the collective
wave functions along the
 collective path is given by the variation of
the overlap between the wave functions corresponding
to different values of $Q_{44}^i$. According to calculations,
the overlaps between the wave functions corresponding to the
SD minimum  and
the extreme hexadecapole deformations considered
($E^*$$\sim$4\,MeV)
vary  between 0.1 and 0.3. These rather large
values
imply the  rigidity of SD states
with respect to the
hexadecapole fields  with $\mu$=4.

A better way of comparing the collectivity along the
collective  paths considered is to perform the
full GCM calculation. This calculation gives at the same time
the distribution  of the lowest wave function as a function
of hexadecapole deformations and the excitation energies
of the associated modes.
We have performed GCM calculations along the three
collective paths defined above. The $Q_{44}^{x}$ path
with a constrained quadrupole deformation has been
used to test the discretization of the hexadecapole
moments, with up to 17 wave functions
corresponding to hexadecapole moments ranging from
4000 to 36000 fm$^4$.
These tests have shown that
eight to ten discretization points lead to
an accuracy of the order of 100\,keV, which is sufficient for
our purpose.

The results for the three collective trajectories are
summarized in Table \ref{T1}.
The HFBCS SD state is predicted at an energy of
1519.41 MeV and its hexadecapole moments
$Q_{44}^z$  and $Q_{44}^x$ are 0
and 20000 fm$^4$, respectively.
These values are not strongly modified by the collective
correlations. The predicted gain in energy
due to dynamical correlations
for the collective
lowest state is in all cases lower than
1.0 MeV, which is much smaller than the typical energy gains
associated with the
quadrupole and  octupole degrees of freedom (see for instance
Ref. \cite {Ska93}). Also the quadrupole and
hexadecapole moments of the first excited
collective states are not
significantly different from the values obtained
for the static minimum. The energies of the
first excited state are large,  $E^*$$>$3\,MeV.
 The
slightly lower value obtained in the unconstrained
$Q_{44}^{x}$ path is probably more due to the variation
of the quadrupole moment along the path than to
a genuine hexadecapole effect.

In Fig.~\ref{FIG3} are shown the squared moduli of
the collective wave functions $g$, Eq.~(\ref{coll}),
 corresponding to
the lowest  GCM state along the
$Q_{44}^{z}$ (top)
and  $Q_{44}^{x}$ (bottom) trajectories
discussed above.
It is seen that the collective ground states do not show any
significant distortion with respect to $Q_{44}$.
The distributions are centered around the static  minimum.
The collective wave function obtained in the  unconstrained $Q_{44}^{x}$
calculations  is slightly asymmetric, i.e.,
components with lower hexadecapole moments
have a slightly larger weight, but this effect is not large
enough to modify the  total hexadecapole moment.

The GCM results that we have discussed show that hexadecapole
correlations are rather weak. In particular, no
hexadecapole excitations
at low energies have been found.
However, the wave functions
considered do not belong to a C$_{4v}$
irreducible representation.
To construct such wave functions, one should consider the mixing
of the four wave functions obtained by rotations of
$\displaystyle{\frac{\pi}{2}}$ around the quantization
axis of the hexadecapole tensor.
For instance, for the C$_4$ symmetry associated with the $z$ axis,
the four corresponding transformations
of the HFBCS wave function ${\vert\Phi\rangle}$ can be
expressed as permutations of coordinates:
\begin{equation}\label{symmetry}
(x,y,z) \Rightarrow (y,-x,z) \Rightarrow (-x,-y,z)  \Rightarrow (-y,x,z).
\end{equation}
 Let us recall that if
one imposes  the condition that the nuclear density is symmetric
with respect to the three planes $x=0$, $y=0$, $z=0$, one can
decompose the single particle wave function $\Phi_k$
into four components with definite parities
with respect to these planes\cite {BON87}. Namely, the wave function
$\Phi_k$ can be written as:
\begin{equation}\label{wf}
\Phi_{k}=
\left(
\begin{array}{c}
\Phi_{k,1}\\
\Phi_{k,2}\\
\Phi_{k,3}\\
\Phi_{k,4}
\end{array}
\right)=
\left(
\begin{array}{c}
Re\Phi_k(\vec r,+)\\
Im\Phi_k(\vec r,+)\\
Re\Phi_k(\vec r,-)\\
Im\Phi_k(\vec r,-)
\end{array}
\right).
\end{equation}
This choice ensures that the parities of the components
$\Phi_{k,\alpha}$ with respect to plane symmetries about
$x$ , $y$ and $z$ planes depend only on the label $\alpha$ and the
parity $p_k$ as indicated in Table \ref{T2}.
Up to an irrelevant phase factor, the permutation
of $x$ and $y$ in  wave function (\ref{wf}) gives rise to
(for positive signature):
\begin{equation}\label{rel2}
\left(
\begin{array}{c}
Re\Phi_k(\vec r,+)\\
Im\Phi_k(\vec r,+)\\
Re\Phi_k(\vec r,-)\\
Im\Phi_k(\vec r,-)
\end{array}
\right)
\Rightarrow
\left(
\begin{array}{c}
-Re\Phi_k(\vec r',+)\\
 Im\Phi_k(\vec r',+)\\
-Im\Phi_k(\vec r',-)\\
-Re\Phi_k(\vec r',-)
\end{array}
\right)
\end{equation}
 where $\vec r'$ is obtained from $\vec r$ by the permutation
of $x$ and $y$. Using relation
(\ref{rel2})
and Table II, one immediately sees
that the second and fourth symmetry operations in
Eq.~(\ref{symmetry})
lead to wave functions with
spin-down component orthogonal to the
 spin-down component of $\Phi_k$, provided
that the $\Phi_{k,\alpha}$  have axial symmetry with
respect to the $z$-axis.
The overlap between the original HFBCS state
and the total wave function obtained by the third transformation
of Eq.~(\ref{symmetry})  is equal to the third component of the
vector density which is exactly zero for time reversal invariant
wave functions.

 The above symmetry properties lead to extremely small overlaps
between the four wave functions generated by the C$_{4v}$
symmetry operations: they are  lower than $10^{-5}$ for all
the wave functions that we have used in our GCM calculation
with a collective path defined along the $z$-axis.
Our calculation has been performed without rotation. Two factors
may increase these overlaps when the nucleus rotates.
The first factor is the change in
the spin content of the wave functions
due to  the Coriolis coupling.
 The second factor is the slight breaking
of axial symmetry due to rotation. The
importance of these effects remains to be studied quantitatively.

\section{CONCLUSIONS}

In this work we have analyzed the presence of hexadecapole
correlations in the
mean fields created by Skyrme like effective interactions.
Hexadecapole deformations have been introduced
by the addition of hexadecapole constraints
aligned either along the symmetry axis or the axis of
rotation.

The static effects of these correlations
turn out to be very weak in SD $^{194}$Hg and do not
seem to be enhanced by rotation. We have also tried to put
into evidence dynamical effects thanks the generator
coordinate method. The GCM ground states obtained
along the three collectives paths introduced do
not show any fingerprint  of polarisation due to the hexadecapole modes.
The first GCM excited modes appear also rather high in energy, at 3-4 MeV.
The lowest energy has been obtained in
a GCM calculation along an $x$-path with relaxed quadrupole moment.
We have interpreted this result as being due mainly to quadrupole
correlations. However, it remains to be verified
in a more complete study
whether hexadecapole
correlations are enhanced by a direct
coupling between   quadrupole and hexadecapole modes.

\begin{center}
{\bf ACKNOWLEDGMENTS}
\end{center}

One of the authors (P.M.) would like to express
his gratitude to the  Service de Physique Nucleaire Th\'eorique
(Universit\'e de Bruxelles) where an important part of this work
has been done, for the warm hospitality.

This work has been supported by the
ARC convention 93/98-166 of the Belgian SSTC,
by the U.S. Department of
Energy through Contracts No. DE-FG05-93ER40770
and DE-AC05-84OR21400,
 and
by the Polish Committee for Scientific
Research under Contract No.~2~P03B~034~08.

\newpage

\begin{figure}[ht]
\caption[FIG1]{
Potential
energy curve as a function of the $Q_{44}^{z}$ moment (top)
and the $Q_{44}^{x}$ moment (bottom).
The solid line represents the result obtained from
particle\--number\--projected wave functions without rotation
and the  dashed line represents the HFBLN results at $I$=42.
The dash-dotted line (bottom) represents the $I$=0
results with a fixed
quadrupole moment.
 } \label{FIG1}
\end{figure}

\begin{figure}[ht]
\caption[FIG2]{%
Mass quadrupole moment, $\langle r^2 Y_{20}^z\rangle$,
versus  $Q_{44}^{x}$
in the nonrotating case (solid line)
and at $I$=42 (dashed line).
}
\label{FIG2}
\end{figure}

\begin{figure}[ht]
\caption[FIG3]{%
Squared moduli of the collective GCM wave functions plotted
versus the $Q_{44}^{z}$ moment (top)
and the $Q_{44}^{x}$ moment (bottom).
Solid and dashed lines represent the results obtained with
unconstrained and constrained quadrupole moments, respectively.
}
\label{FIG3}
\end{figure}

\begin{table}
\begin{tabular}{c|ccc|ccc|ccc}
 $n$ & $Q_{44}^{z}$ & Energy &$e^{*}$
     & $Q_{44}^{x}$ & Energy &$e^{*}$
     & $Q_{44}^{x c}$ & Energy &$e^{*}$ \\
\hline
 1 &    3.5   &-1520.12  & 0.00
   &   21.2   &-1520.15  & 0.00
   &   20.4   &-1520.36  & 0.00   \\
 2 &  -27.0   &-1516.10  & 4.02
   &   20.1   &-1516.95  & 3.21
   &   23.6   &-1516.17  & 4.19   \\
 3 &   31.9   &-1514.47  & 5.65
   &   15.5   &-1515.19  & 4.97
   &   24.1   &-1515.14  & 5.22   \\
\end{tabular}
\caption{The GCM results for the
 three collective paths specified by $Q_{44}^{z}$,
 $Q_{44}^{x}$ and $Q_{44}^{x c}$, where
 superscript $c$ denotes the collective path with constrained quadrupole
 moment.
  Excitation energies are denoted by $e^{*}$.
  Energies are given in $MeV$ units and hexadecapole moments in
  fm$^{4}$ for the $Q_{44}^{z}$ path and in 1000 fm$^4$
 in the two other cases. } \label{T1}
\end{table}

\narrowtext
\begin{table}
\begin{tabular}{c|ccc}
$\alpha$ & $x$ & $y$ & $z$ \\
\tableline
1 & $+$ & $+$ & $p_k$ \\ \hline
2 & $-$ & $-$ & $p_k$ \\ \hline
3 & $-$ & $+$ & $-p_k$ \\ \hline
4 & $+$ & $-$ & $-p_k$
\end{tabular}
\caption{\label{T2}Parities of the components $\Phi_{k,\alpha}$ of a
function $\Phi_k$ of parity $p_k$, with respect to the
$x=0$, $y=0$ and $z=0$ planes.}
\end{table}

\end{document}